\documentclass[seceq]{ptptex}
\usepackage{wrapft}
\usepackage{graphicx}
\usepackage{amsmath}
\usepackage{bm}

\begin{document}
\markboth{
T.~Tatekawa
}{
4LPT for cosmological dust fluid}

\title{
Fourth-order perturbative equations in Lagrangian
perturbation theory for a cosmological dust fluid
}

\author{
Takayuki \textsc{Tatekawa}$^{1,2}$
\footnote{E-mail: tatekawa@u-fukui.ac.jp}
}

\inst{
$^1$Center for Information Initiative, University of Fukui,
3-9-1 Bunkyo,  Fukui 910-8507, Japan
\\
$^2$Research Institute for Science and Engineering, Waseda University,
3-4-1 Okubo, Shinjuku-ku, Tokyo 169-8555, Japan
}



\abst{
We have derived fourth-order perturbative equations in
Lagrangian perturbation theory for a cosmological dust fluid.
These equations are derived under the supposition of Newtonian cosmology
in the Friedmann-Lema\^{i}tre-Robertson-Walker Universe model.
Even if we consider the longitudinal mode in the first-order
perturbation, the transverse mode appears in the
third-order perturbation. Furthermore, in this
case, six longitudinal-mode equations and four transverse-mode
equations appear in the fourth-order perturbation.
The application of the fourth-order perturbation leads to
a precise prediction of the large-scale structure.
}

\PTPindex{461, 469}   

\maketitle

\section{Introduction}\label{sec:intro}

Describing the large-scale structure in the Universe is one of
the most important problems in modern cosmology. It has
been postulated that this structure was formed from the
primordial density fluctuation due to
self-gravitational instability
~\cite{Peebles1980,Padmanabhan1993,Sahni1995,Coles1995,Bernardeau2002,Jones2004,Tatekawa2005,Tatekawa2010}. Given that the formation of this structure depends
upon cosmic expansion, models of the dark energy
in the Universe would be constrained by the observation
of the large-scale structure.

One theoretical prediction for the formation of the structure in the Universe,
nonlinear perturbation theory, has been considered for
a long time. In particular, we notice Lagrangian perturbation
of the cosmological fluid in Newtonian cosmology.
The behavior of the cosmological fluid is constrained by the assumption of a homogeneous and isotropic Universe in the
model (e.g., the Friedmann-Lema\^{i}tre-Robertson-Walker Universe model).
The cosmic expansion is given by Friedmann equations or similar
equations.
Zel'dovich proposed a model of the cosmological fluid based on Lagrangian perturbation
\cite{Zeldovich1970}. Although the perturbation remains
in the linear region, density fluctuation in the quasilinear
region can be described~\cite{Zeldovich1970,Arnold1982,Shandarin1989}.
Once these descriptions are available, higher-order perturbative solutions can be
derived. Perturbative equations up to the third order describing
dust (pressureless fluid) and a polytropic fluid have previously been developed
\cite{Bouchet1992,Buchert1993,Buchert1994,Bouchet1995,Catelan1995,Buchert1992,Barrow1993,Sasaki1998,Morita01,Tatekawa02,Tatekawa05A,Tatekawa05B}.
An improved method was subsequently proposed~\cite{Yoshisato1998,Tatekawa2007}, in which the procedure for determining the expansion of the
perturbation was changed; however, a significant improvement was not seen.

Recently, resummation theories have been proposed~\cite{Crocce06A,Taruya08,Bernardeau08B,Matsubara08,Bernardeau08}; these theories reorganize an infinite series of diagrams
into the form of standard perturbation theory and partially resum them.
Resummation theories can precisely describe physical quantities such as
power spectra and correlation functions. In particular,
theories based on Lagrangian perturbation have several merits:
they make is easy to describe physical quantities in not only real space
but also in redshift space
\cite{Matsubara08,Matsubara11,Okamura11} and
nonlocal bias is included~\cite{Matsubara08B,Matsubara11}.
A formula for arbitrary-order ($m$-loop) correction has
been proposed~\cite{Okamura11}.
If we have greater knowledge of higher-order Lagrangian perturbations,
we can further improve resummation theories that are based on Lagrangian
perturbation. For example, recently, the matter bispectrum
with higher-order Lagrangian perturbations was computed\cite{Rampf12A, Rampf12B}.

Although perturbation theory cannot describe strongly
nonlinear regions, it still plays an important role.
Cosmological $N$-body simulations have been applied to describe
the evolution of strongly nonlinear regions. The initial condition
for the simulations is given by Lagrangian perturbation.
It is known that the accuracy of the initial condition
affects time evolution in strongly nonlinear regions~\cite{Scoccimarro1998,Crocce06B,Tatekawa07C,Jenkins2009}.
In the future, this knowledge would be important for precision cosmology.

The application of  Lagrangian perturbation theory (LPT) to observational
results will also become important.
For example, the Sloan Digital Sky Survey (SDSS) Data Release 7 (DR7)
contains spectroscopic data on approximately one million galaxies
\cite{SDSS-DR7}.
In other words, the SDSS project
provides a three-dimensional map of the large-scale structure.
Because the magnitude limit of the project is approximately $22$,
the survey is carried out within the low-$z$ region.
In future surveys, high-$z$ ($z>1$) galaxies are expected to
be observed. It is expected that the results of these surveys will provide us with knowledge on the evolution of the large-scale
structure. Furthermore, because the high-$z$ region is still quasi-nonlinear, higher-order LPT will be valid.
We expect that very precise predictions regarding the large-scale structure
can be made by incorporating fourth-order perturbation into
resummation theory.

This paper is organized as follows. In Sec.~\ref{sec:basic}, we briefly
introduce the basic equations for Lagrangian perturbation; in this study, we consider only a dust fluid.
Next, we present the perturbative equations in Sec.~\ref{sec:Perturbation}.
The fourth-order perturbative equations consist of
six longitudinal-mode (scalar) equations and four transverse-mode
(vector) equations. We also derive the perturbative solutions
in the Einstein-de Sitter (E-dS) Universe model.
Every mode grows at the same rate; thus, it is not possible to ignore any
part of these modes.
Next, the application of these equations to our results is considered.
In Sec.~\ref{sec:appli}, we discuss the applications of the higher-order
perturbations.
In Sec.~\ref{subsec:resum}, we describe one possible outcome of applying
our result to resummation theories.
In Sec.~\ref{subsec:init}, we consider initial condition problems
in cosmological $N$-body simulations.
Finally, we summarize our findings in Sec.~\ref{sec:summary}.
In Appendix \ref{sec:det}, we describe the relation between the determinant
of the displacement matrix and the combination of triplet terms.
Appendix \ref{sec:dev-4th} presents a detailed derivation of the fourth-order
perturbative equations. Appendix \ref{sec:loop-p} presents higher-order
loop correction for the power spectrum.

\section{Basic equations}\label{sec:basic}
In this section, we briefly introduce the Lagrangian description of a Newtonian
cosmological fluid. When the scale of objects is smaller than
that of the horizon, this description is valid. The effect of
cosmic expansion is reflected by the scale factor $a$.
The solution $a$ is given by the Friedmann equations (or alternative
equations).
In the comoving coordinates,
the basic equations for a cosmological dust fluid are %
\begin{eqnarray}
\frac{\partial \delta}{\partial t} + \frac{1}{a} \nabla_x \cdot \{ \bm{v}
(1+\delta) \} &=& 0 , \label{eqn:comoving-conti-eq} \\
\frac{\partial \bm{v}}{\partial t} + \frac{1}{a} (\bm{v} \cdot \nabla_x)
\bm{v} + \frac{\dot{a}}{a} \bm{v} &=& \frac{1}{a} \tilde{\bm{g}}
 , \label{eqn:comoving-Euler-eq} \\
\nabla_x \times \tilde{\bm{g}} &=& \bm{0} , \label{eqn:rot-g} \\
\nabla_x \cdot \tilde{\bm{g}} &=& - 4 \pi G \rho_b a \delta ,
\label{eqn:comoving-Poisson-eq} \\
\delta & \equiv & \frac{\rho - \rho_b}{\rho_b} \,.
\end{eqnarray}
In Eulerian perturbation theory, the density fluctuation $\delta$
is regarded as a perturbative quantity. However, in
LPT, displacement from a homogeneous
distribution is considered:
\begin{equation} \label{eqn:x=q+s}
\bm{x} = \bm{q} + \bm{s} (\bm{q},t) ,
\end{equation}
where $\bm{x}$ and $\bm{q}$ are the comoving Eulerian coordinates
and the Lagrangian coordinates, respectively. $\bm{s}$ is
the displacement vector, which is regarded as a perturbative quantity.
From Eq.~(\ref{eqn:x=q+s}), we can solve the continuous equation
~(\ref{eqn:comoving-conti-eq}) exactly. Then, the density
fluctuation is given in the formally exact form
\begin{equation} \label{eqn:exact-delta}
\delta = 1 - J^{-1}, ~~ J \equiv \det \left (
 \frac{\partial x_i}{\partial q_j} \right ) \,.
\end{equation}
$J$ refers to the Jacobian of the coordinate transformation
from Eulerian $\bm{x}$ to Lagrangian $\bm{q}$.
Therefore, when we derive the solutions for $\bm{s}$, we can determine
the evolution of the density fluctuation.

The peculiar velocity is given as follows:%
\begin{equation}
\bm{v}=a \dot{\bm{s}} \label{eqn:L-velocity} .
\end{equation}
Then, we introduce the Lagrangian time derivative
\begin{equation}
\frac{\rm d}{{\rm d} t} \equiv \frac{\partial}{\partial t}
 + \frac{1}{a} \bm{v} \cdot \nabla_x . \label{eqn:dt-L}
\end{equation}
Taking the divergence and
rotation of Eq.~(\ref{eqn:comoving-Euler-eq}), we obtain
the evolution equations for the Lagrangian displacement:
\begin{eqnarray}
\nabla_x
 \cdot \left (\ddot{\bm{s}}
  + 2 \frac{\dot{a}}{a} \dot{\bm{s}} \right )
&=& -4 \pi G \rho_b (J^{-1} -1) , \label{eqn:L-longi-eqn} \\
\nabla_x \times \left (\ddot{\bm{s}} + 2 \frac{\dot{a}}{a}
 \dot{\bm{s}} \right )
&=& 0 , \label{eqn:L-trans-eqn}
\end{eqnarray}
where $(\dot{})$ refers to the Lagrangian time derivative (Eq.~(\ref{eqn:dt-L})).
To solve the perturbative equations,
we decompose the Lagrangian perturbation into its longitudinal
and transverse modes:
\begin{eqnarray}
\bm{s} &=& \nabla S + \bm{s}^T , \\
\nabla \cdot \bm{s}^T &=& 0 ,
\end{eqnarray}
where $\nabla$ refers to the Lagrangian spatial derivative.

Here, we expand the Jacobian:
\begin{eqnarray} \label{eqn:J-expansion}
J &=& 1 + \nabla^2 S + \frac{1}{2} \left \{ (\nabla^2 S)^2
  - S_{,ij} S_{,ji} - s^T_{i,j} s^T_{j,i}
  - 2 S_{,ij} s^T_{j,i} \right \} \nonumber \\
  &&
  + \det \left (S_{,ij} + s^T_{i,j} \right ) .
\end{eqnarray}
Eqs.~(\ref{eqn:L-longi-eqn}) and (\ref{eqn:L-trans-eqn})
include the Eulerian spatial derivative, and we convert it to the Lagrangian
spatial derivative:
\begin{eqnarray*}
\frac{\partial}{\partial x_i} &=& \frac{\partial}{\partial q_i}
 - s_{j,i} \frac{\partial}{\partial x_j} \\
 &=& \frac{\partial}{\partial q_i}
 - s_{j,i} \frac{\partial}{\partial q_j}
 + s_{j,i} s_{k,j} \frac{\partial}{\partial x_k} \\
 &=& \frac{\partial}{\partial q_i}
 - s_{j,i} \frac{\partial}{\partial q_j}
 + s_{j,i} s_{k,j} \frac{\partial}{\partial q_k} + \cdots .
\end{eqnarray*}

\section{Lagrangian perturbative equations}
\label{sec:Perturbation}

\subsection{First-order perturbative equations}
\label{subsec:1st-eq}

The first-order perturbative equation in the longitudinal mode
was derived by Zel'dovich~\cite{Zeldovich1970}.
The evolution equation is given as follows:
\begin{equation}
\ddot{S}^{(1)} + 2 \frac{\dot{a}}{a}
 \dot{S}^{(1)} - 4 \pi G \rho_b S^{(1)} = 0 .
 \label{eqn:1st-L-eq}
\end{equation}
The solution can be divided into its spatial and temporal parts:
\begin{eqnarray}
S^{(1)} &=& g_1 (t) \psi^{(1)} (\bm{q}) , \label{eqn:1st-L-st}
\\
\ddot{g}_1 + 2 \frac{\dot{a}}{a} \dot{g}_1
 - 4 \pi G \rho_b g_1 &=& 0 \label{eqn:1st-L-t-eq} .
\end{eqnarray}
In the E-dS Universe model ($a(t) \propto t^{2/3}$), the solutions to
the temporal part are given as
\begin{equation}
g_1(t) \propto t^{2/3}, t^{-1} .
\end{equation}
Hereafter, we consider only the growing solution ($g_1(t) = t^{2/3}$).
The spatial part $\psi^{(1)} (\bm{q})$ is given by the initial condition.

The effect of the transverse mode was examined by
Buchert~\cite{Buchert1992} and Barrow and Saich~\cite{Barrow1993}:
\begin{equation}
\nabla \times \left (\ddot{\bm{s}}^{T (1)} +
 2 \frac{\dot{a}}{a} \dot{\bm{s}}^{T (1)} \right )
 = \bm{0} .
\end{equation}
We choose a reasonable boundary condition for this equation.
The spatial differential operator can be removed, leaving
\begin{equation}
\ddot{\bm{s}}^{T (1)} +
 2 \frac{\dot{a}}{a} \dot{\bm{s}}^{T (1)} = \bm{0} .
\end{equation}
In the E-dS Universe model, the solution is given as
\begin{equation}
\bm{s}^{T (1)} \propto t^0, t^{-1/3} .
\end{equation}
Because there are no growing solutions for first-order perturbation
in the transverse mode, we consider only the longitudinal mode in the
first-order perturbation.

\subsection{Second-order perturbative equations}
\label{subsec:2nd-eq}

The second-order perturbative equation in the longitudinal
mode was derived by
Bouchet \textit{et al.}~\cite{Bouchet1992} and
Buchert and Ehlers~\cite{Buchert1993}.
The evolution equation for the
second-order longitudinal mode is derived from Eq.~(\ref{eqn:L-longi-eqn}).
This equation is given as follows:
\begin{equation}
\ddot{S}_{,ii}^{(2)} + 2 \frac{\dot{a}}{a}
 \dot{S}_{,ii}^{(2)} - 4 \pi G \rho_b S_{,ii}^{(2)}
 = -2 \pi G \rho_b \left ( S_{,ii}^{(1)} S_{,jj}^{(1)}
 - S_{,ij}^{(1)} S_{,ij}^{(1)} \right ) .
\end{equation}
We decompose the perturbation into its spatial and
temporal parts:
\begin{equation} \label{eqn:2nd-L-ts}
S^{(2)} = g_2(t) \psi^{(2)} (\bm{q}) .
\end{equation}
The evolution equation is then given as follows:
\begin{eqnarray}
\ddot{g}_2 + 2 \frac{\dot{a}}{a} \dot{g}_2
 - 4 \pi G \rho_b g_2 &=& - 4 \pi G \rho_b g_1^2 ,
 \label{eqn:2nd-L-t-eq} \\
\psi_{,ii}^{(2)} &=& \frac{1}{2} \left \{
 \psi_{,ii}^{(1)} \psi_{,jj}^{(1)} - \psi_{,ij}^{(1)}
 \psi_{,ij}^{(1)} \right \} . \label{eqn:2nd-L-spa-eq}
\end{eqnarray}
In the E-dS Universe model, the special solution for the temporal part
is given as follows:
\begin{equation}
g_2(t) = -\frac{3}{7} t^{4/3} .
\end{equation}

The equation in the transverse mode is given as follows:
\begin{equation}
\nabla \times \left (\ddot{\bm{s}}^{T (2)} +
 2 \frac{\dot{a}}{a} \dot{\bm{s}}^{T (2)} \right )
 = \bm{0} .
\end{equation}
The evolution
equation for the second-order perturbation coincides with that of the
first-order perturbation.
In other words, there are no perturbative solutions
for the second-order transverse mode.

\subsection{Third-order perturbative equations}
\label{subsec:3rd-eq}

The third-order perturbative solutions in the longitudinal mode were derived
by Buchert~\cite{Buchert1994}, Bouchet \textit{et al.}~\cite{Bouchet1995},
and Catelan~\cite{Catelan1995}. The evolution equation for the
third-order longitudinal mode is derived from Eq.~(\ref{eqn:L-longi-eqn}):

\begin{eqnarray}
&& \ddot{S}_{,ii}^{(3)} + 2 \frac{\dot{a}}{a} \dot{S}_{,ii}^{(3)}
 - 4 \pi G \rho_b S_{,ii}^{(3)} \nonumber \\
&=& 4 \pi G \rho_b \left [ -g_1 (g_2 - g_1^2)
 \left \{ \psi_{,ii}^{(1)} \psi_{,jj}^{(2)} - \psi_{,ij}^{(1)} \psi_{,ij}^{(2)}
 \right \} - 2 g_1^3 \det \left ( \psi_{,ij}^{(1)} \right ) \right ]
 .
\end{eqnarray}
Here, we substitute the first- and second-order equations,
(\ref{eqn:1st-L-eq})--(\ref{eqn:1st-L-t-eq}) and
(\ref{eqn:2nd-L-ts})--(\ref{eqn:2nd-L-spa-eq}), respectively, into the third-order equation.

The transform between the determinant and the triplet terms is described
in Appendix \ref{sec:det}. We decompose the perturbation into two spatial and two
temporal parts:
\begin{equation}
S^{(3)} = g_{3a}(t) \psi^{(3a)} (\bm{q}) + g_{3b}(t) \psi^{(3b)} (\bm{q}) .
\end{equation}
The evolution equations are given as follows:
\begin{eqnarray}
\ddot{g}_{3a} + 2 \frac{\dot{a}}{a} \dot{g}_{3a}
 - 4 \pi G \rho_b g_{3a} &=& - 8 \pi G \rho_b g_1 (g_2 - g_1^2 ) ,
 \label{eqn:3rd-L-t-eqA} \\
\ddot{g}_{3b} + 2 \frac{\dot{a}}{a} \dot{g}_{3b}
 - 4 \pi G \rho_b g_{3b} &=& - 8 \pi G \rho_b g_1^3 ,
 \label{eqn:3rd-L-t-eqB}
\end{eqnarray}
\begin{eqnarray}
\psi_{,ii}^{(3a)} &=& \frac{1}{2} \left \{
 \psi_{,ii}^{(1)} \psi_{,jj}^{(2)} - \psi_{,ij}^{(1)}
 \psi_{,ij}^{(2)} \right \} , \label{eqn:3rd-L-spa-eqA} \\
\psi_{,ii}^{(3b)} &=& \det \left ( \psi_{,ij}^{(1)} \right ) \nonumber \\
 &=& \frac{1}{6} \psi_{,ii}^{(1)} \psi_{,jj}^{(1)} \psi_{,kk}^{(1)}
 - \frac{1}{2} \psi_{,ii}^{(1)} \psi_{,jk}^{(1)} \psi_{,jk}^{(1)}
 + \frac{1}{3} \psi_{,ij}^{(1)} \psi_{,jk}^{(1)} \psi_{,ki}^{(1)} .
 \label{eqn:3rd-L-spa-eqB}
\end{eqnarray}
In the E-dS Universe model, the special solutions to the temporal parts are given as follows:
\begin{eqnarray}
g_{3a}(t) &=& \frac{10}{21} t^2 , \\
g_{3b}(t) &=& -\frac{1}{3} t^2 .
\end{eqnarray}

Sasaki and Kasai~\cite{Sasaki1998} derived
third-order perturbative solutions to both the longitudinal
and transverse modes. In this study, we only consider
the effect of the first-order longitudinal growing mode.
The evolution equation for the
third-order transverse mode is derived from Eq.~(\ref{eqn:L-trans-eqn}).
By using first- and second-order longitudinal-mode equations,
(\ref{eqn:1st-L-eq})--(\ref{eqn:1st-L-t-eq}) and
(\ref{eqn:2nd-L-ts})--(\ref{eqn:2nd-L-spa-eq}), respectively,
the equation for the third-order transverse mode is given as follows:
\begin{equation}
 \varepsilon_{ijk}
 \left (\ddot{s}_{k,j}^{T (3)} + 2 \frac{\dot{a}}{a}
 \dot{s}_{k,j}^{T (3)} \right ) - g_1 (g_2 - g_1^2) \varepsilon_{ijk}
 \psi_{,jl}^{(1)} \psi_{,kl}^{(2)}
- g_1 g_2 \varepsilon_{ijk}
 \psi_{,jl}^{(2)} \psi_{,kl}^{(1)} = 0 .
\end{equation}
We decompose the perturbation into its spatial and temporal parts:
\begin{equation} \label{eqn:3rd-T-ts}
s_i^{T (3)} = g_{3T} (t) \zeta_i^{(3)} (\bm{q}) .
\end{equation}
The evolution equations are given as follows:
\begin{eqnarray}
\ddot{g}_{3T} + 2 \frac{\dot{a}}{a}
\dot{g}_{3T} &=& 4 \pi G \rho_b g_1^3 , \\
- \nabla^2 \zeta_i^{(3)} &=& \left ( \psi_{,il}^{(1)} \psi_{,kl}^{(2)}
 - \psi_{,kl}^{(1)} \psi_{,il}^{(2)} \right )_{,k} .
\end{eqnarray}
Even if we consider only the longitudinal mode in the first-order perturbation,
a third-order perturbation appears in the transverse mode.
The meaning of the transverse mode will be examined in Sec.~\ref{sec:summary}.

In the E-dS Universe model, the solution for the temporal part is given as follows:
\begin{equation}
g_{3T}(t) = \frac{1}{7} t^2 .
\end{equation}
%
\subsection{Fourth-order perturbative equations}
\label{subsec:4th-eq}
Although fourth-order perturbative solutions have been derived in the past
(for example, ~\cite{Vanselow1995, Shiraishi1995}), these solutions were reported before the publication of \cite{Sasaki1998}.
Therefore, these derivations ignored the effect of transverse modes.
In this paper, we consider the transverse modes.

The fourth-order longitudinal mode is also affected by
the third-order transverse mode, which is generated
by the first- and second-order longitudinal modes.

The derivation of the fourth-order perturbative equations
is presented in Appendix~\ref{sec:dev-4th}.
The fourth-order perturbative equations for
the longitudinal mode can be derived.
In this section, we decompose the perturbation into twelve different parts, six spatial and six temporal:
\begin{eqnarray}
S^{(4)} &=& g_{4a}(t) \psi^{(4a)} (\bm{q}) + g_{4b}(t) \psi^{(4b)} (\bm{q})
 + g_{4c}(t) \psi^{(4c)} (\bm{q}) \nonumber \\
&& + g_{4d}(t) \psi^{(4d)} (\bm{q})
 + g_{4e}(t) \psi^{(4e)} (\bm{q}) + g_{4f}(t) \psi^{(4f)} (\bm{q}) .
\end{eqnarray}
The evolution equations are given as follows:
\begin{eqnarray}
\ddot{g}_{4a} + 2 \frac{\dot{a}}{a} \dot{g}_{4a}
 - 4 \pi G \rho_b g_{4a} &=& - 8 \pi G \rho_b g_1 ( g_{3a} - 2g_1 g_2 + 2 g_1^3 ) , \label{eqn:4th-L-g4a} \\
\ddot{g}_{4b} + 2 \frac{\dot{a}}{a} \dot{g}_{4b}
 - 4 \pi G \rho_b g_{4b} &=& - 8 \pi G \rho_b g_1 ( g_{3b} - 2 g_1^3 ) ,
 \label{eqn:4th-L-g4b} \\
\ddot{g}_{4c} + 2 \frac{\dot{a}}{a} \dot{g}_{4c}
 - 4 \pi G \rho_b g_{4c} &=& 4 \pi G \rho_b g_1^4 ,
 \label{eqn:4th-L-g4c} \\
\ddot{g}_{4d} + 2 \frac{\dot{a}}{a} \dot{g}_{4d}
 - 4 \pi G \rho_b g_{4d} &=& -4 \pi G \rho_b g_2 (g_2 - 2 g_1^2) ,
 \label{eqn:4th-L-g4d} \\
\ddot{g}_{4e} + 2 \frac{\dot{a}}{a} \dot{g}_{4e}
 - 4 \pi G \rho_b g_{4e} &=& 4 \pi G \rho_b g_1^2 (2 g_2 - g_1^2) ,
 \label{eqn:4th-L-g4e} \\
\ddot{g}_{4f} + 2 \frac{\dot{a}}{a} \dot{g}_{4f}
 - 4 \pi G \rho_b g_{4f} &=& 4 \pi G \rho_b g_1^4 ,
 \label{eqn:4th-L-g4f}
\end{eqnarray}
\begin{eqnarray}
\psi_{,ii}^{(4a)} &=& \frac{1}{2}
 \left ( \psi_{,ii}^{(1)} \psi_{,jj}^{(3a)} - \psi_{,ij}^{(1)}
 \psi_{,ij}^{(3a)} \right ) , \label{eqn:4th-L-psi4a} \\
\psi_{,ii}^{(4b)} &=& \frac{1}{2}
 \left ( \psi_{,ii}^{(1)} \psi_{,jj}^{(3b)} - \psi_{,ij}^{(1)}
 \psi_{,ij}^{(3b)} \right ) , \label{eqn:4th-L-psi4b} \\
\psi_{,ii}^{(4c)} &=& \psi_{,ij}^{(1)} \zeta_{i,j}^{(3)} ,
 \label{eqn:4th-L-psi4c} \\
\psi_{,ii}^{(4d)} &=& \frac{1}{2}
 \left ( \psi_{,ii}^{(2)} \psi_{,jj}^{(2)}
 - \psi_{,ij}^{(2)} \psi_{,ij}^{(2)} \right ) ,
 \label{eqn:4th-L-psi4d} \\
\psi_{,ii}^{(4e)} &=& \psi_{,ii}^{(1)} \psi_{,jk}^{(1)} \psi_{,jk}^{(2)}
 - \psi_{,ij}^{(1)} \psi_{,jk}^{(1)} \psi_{,ki}^{(2)}
 - \frac{3}{8} \psi_{,ii}^{(2)}
  \left \{ \psi_{,jj}^{(1)} \psi_{,kk}^{(1)}
 - \psi_{,jk}^{(1)} \psi_{,jk}^{(1)} \right \} ,
 \label{eqn:4th-L-psi4e} \\
\psi_{,ii}^{(4f)} &=& - \frac{7}{16} \left \{ \psi_{,ii}^{(1)} \psi_{,jj}^{(1)}
   - \psi_{,ij}^{(1)} \psi_{,ij}^{(1)} \right \}
  \left \{ \psi_{,kk}^{(1)} \psi_{,ll}^{(1)}
   - \psi_{,kl}^{(1)} \psi_{,kl}^{(1)} \right \} \nonumber \\
&&  + \frac{7}{3} \psi_{,ii}^{(1)} \psi_{,jj}^{(1)}
  \psi_{,kk}^{(1)} \psi_{,ll}^{(1)} -
 \frac{4}{3} \psi_{,ii}^{(1)} \psi_{,jk}^{(1)} \psi_{,kl}^{(1)}
 \psi_{,li}^{(1)}
 + \psi_{,ij}^{(1)} \psi_{,jk}^{(1)} \psi_{,kl}^{(1)} \psi_{,li}^{(1)}
 . \label{eqn:4th-L-psi4f}
\end{eqnarray}
In the E-dS Universe model, the special solutions to the temporal parts are given as follows:
\begin{eqnarray}
g_{4a}(t) &=& -\frac{20}{33} t^{8/3} , \\
g_{4b}(t) &=& \frac{14}{33} t^{8/3} , \\
g_{4c}(t) &=& \frac{1}{11} t^{8/3} , \\
g_{4d}(t) &=& -\frac{51}{539} t^{8/3} , \\
g_{4e}(t) &=& -\frac{13}{77} t^{8/3} , \\
g_{4f}(t) &=& \frac{1}{11} t^{8/3} .
\end{eqnarray}

Next, we consider the evolution equation in the transverse mode.
The evolution equation for the
fourth-order transverse mode is derived from Eq.~(\ref{eqn:L-trans-eqn}).
We can decompose the perturbation into eight parts, four spatial and
four temporal:
\begin{equation} \label{eqn:4th-T-decompose}
s_k^{T (4)} = g_{4a}(t) \zeta_k^{(4a)} (\bm{q}) + g_{4b}(t) \zeta_k^{(4b)} (\bm{q})
 + g_{4c}(t) \zeta_k^{(4c)} (\bm{q}) + g_{4d}(t) \zeta_k^{(4d)} (\bm{q}) .
\end{equation}
The fourth-order perturbative equations for the transverse mode
are decomposed as follows:
\begin{eqnarray}
\ddot{g}_{4Ta} + 2 \frac{\dot{a}}{a} \dot{g}_{4Ta} &=& -8 \pi G \rho_b g_1^2
 (g_2- g_1^2 ) , \label{eqn:4th-T-g4a} \\
\ddot{g}_{4Tb} + 2 \frac{\dot{a}}{a} \dot{g}_{4Tb} &=& -8 \pi G \rho_b g_1^4
 , \label{eqn:4th-T-g4b} \\
\ddot{g}_{4Tc} + 2 \frac{\dot{a}}{a} \dot{g}_{4Tc} &=& 4 \pi G \rho_b g_1
 (g_{3T} + g_1^3) , \label{eqn:4th-T-g4c} \\
\ddot{g}_{4Td} + 2 \frac{\dot{a}}{a} \dot{g}_{4Td} &=& -4 \pi G \rho_b g_1^4
 , \label{eqn:4th-T-g4d}
\end{eqnarray}
\begin{eqnarray}
- \nabla^2 \zeta_i^{(4a)} &=& \left ( \psi_{,il}^{(1)} \psi_{,kl}^{(3a)}
 - \psi_{,kl}^{(1)} \psi_{,il}^{(3a)} \right )_{,k}
 , \label{eqn:4th-T-psi4a} \\
- \nabla^2 \zeta_i^{(4b)} &=& \left ( \psi_{,il}^{(1)} \psi_{,kl}^{(3b)}
 - \psi_{,kl}^{(1)} \psi_{,il}^{(3b)} \right )_{,k}
 , \label{eqn:4th-T-psi4b} \\
- \nabla^2 \zeta_i^{(4c)} &=& \left ( \psi_{,il}^{(1)} \zeta_{k,l}^{(3)}
 - \psi_{,kl}^{(1)} \zeta_{i,l}^{(3)} \right )_{,k}
 , \label{eqn:4th-T-psi4c} \\
- \nabla^2 \zeta_i^{(4d)} &=& \left ( \psi_{,ij}^{(1)} \psi_{,jk}^{(1)}
 \psi_{,kl}^{(2)} - \psi_{,jk}^{(1)} \psi_{,kl}^{(1)} \psi_{,ij}^{(2)}
  \right )_{,l} . \label{eqn:4th-T-psi4d}
\end{eqnarray}
In the E-dS Universe model, the special solutions to the temporal parts are given as follows:
\begin{eqnarray}
g_{4Ta}(t) &=& \frac{5}{21} t^{8/3} , \\
g_{4Tb}(t) &=& -\frac{1}{6} t^{8/3} , \\
g_{4Tc}(t) &=& \frac{2}{21} t^{8/3} , \\
g_{4Td}(t) &=& -\frac{1}{12} t^{8/3} .
\end{eqnarray}

The fourth-order perturbative equations in this subsection are the main result
of this paper. In E-dS Universe model, $g_4$ is proportional to $a^4$.

For the generic Friedmann Universe model, we must solve the equations for the
temporal parts. According to the behavior of $g_2$ and $g_3$,
$g_4$ would be proportional to $a^4$ mostly.
In the matter-dominant era, the solutions almost
coincide with those in E-dS Universe model.

\section{Implication of fourth-order solutions for practical
applications}
\label{sec:appli}

\subsection{Applications to resummation theory}
\label{subsec:resum}

Although the precision of the perturbation improves with increasing order,
the effect of higher-order perturbation is quite small.
When the density fluctuation evolves in the nonlinear region,
it is difficult to describe the evolution using perturbation theory.
Recently,
a set of remarkable methods that substantially improve the precision
has been studied. These methods constitute the resummation theory~\cite{Crocce06A,Taruya08,Bernardeau08B,Matsubara08}.

There are many methods of the resummation theory, and still new methods
have been also developed. Hiramatsu and Taruya~\cite{Hiramatsu2009}
have developed numerical algorithm to solve closure evolutions in both 
perturbative and nonperturbative regimes. For time evolution of matter
power spectra, the calculation result in nonperturbative regime almost
coincides to that from $N$-body simulations. 
Crocce, Scoccimarro and Bernardeau~\cite{Crocce2012} have developed
fast renormalized perturbative scheme. The feature of the scheme is to
apply multi-point propagators.  They can derive nonlinear power spectra quickly. 
Pietroni~\cite{Pietroni2008} proposed a different method. The method is
similar to the familiar BBGKY hierarchy~\cite{Peebles1980}. 
To solve group of evolution equations, we must introduce some kind of truncation. 
In this paper. as one assumption, a part of four-point function is truncated. In
other words, four-point correlator is consist from two-point correlators.
Of course, these works consider generic Friedmann Universe models.
In these works, although the growth rate of $n$-th order perturbation is
given by $g_1^n$, the power spectrum is calculated with high precision.

As we mentioned in Sec.~\ref{sec:basic}, the relation between the density
fluctuation and the Lagrangian perturbation is nonlinear. In other words,
Lagrangian linear perturbation includes a nonlinear effect in the
density fluctuation (Eulerian perturbation). Therefore, if we choose
Lagrangian perturbation as the model, a subset of the infinite series of diagrams
in Eulerian perturbation would be included~\cite{Bernardeau08B}.
Furthermore, the resummation based on LPT can compute the power spectrum
in redshift space. The mapping from real space and redshift space is given
analytically.

Recently, Okamura \textit{et al.}~\cite{Okamura11}
proposed a generic formula for
$m$-loop resummation via LPT.
The formula is shown in Appendix \ref{sec:loop-p}.
Their formula can apply to generic Friedmann Universe models.
They described the evolution of baryon acoustic oscillation (BAO)~\cite{Eisenstein98,Meiksin99}.
Additionally, they compared improved LPT to $N$-body simulations in real space.
Their two-loop correction can predict the power spectrum with a higher
accuracy than when using a one-loop correction. However, because one-loop
correction is sufficiently accurate
for describing the nonlinearity, the effect of the two-loop correction is
shown to be less significant for two-point correlation functions.

Their generic formula is applied to not only the power spectrum
in real space but also in redshift space. Furthermore, their formula
is adopted in the generic Friedmann Universe model. In this paper, we show
the apparent form of the perturbative solutions in a flat Universe model
without a cosmological constant.

When we consider the transverse mode in third-order LPT (3LPT), 
at least the power spectrum
in redshift space would change. The transverse mode in 3LPT does not
affect the growth of the density fluctuation (\ref{eqn:exact-delta}).
However, the displacement whose direction is different from that of
the longitudinal mode is given. Details of the difference between
the directions of the longitudinal and transverse modes
will be discussed in next subsection.

For application of the resummation theory, higher-order perturbative
solutions are the point. For two-loop correction for the power spectrum,
we require the perturbative solutions up to 3LPT.
For more  improvement of the power spectrum, we will consider three-loop
correction. As we show in Appendix \ref{sec:loop-p}, 5LPT is required
for three-loop correction. 

Making a comparison between one- or
two-loop corrections and $N$-body simulations, one finds that two-loop LPT
is valid for $k \le 0.3 h^{-1} \,\mbox{Mpc}^{-1}$ at $z=2$ with
1\% accuracy.

On the other hand, when we consider bispectrum, 4LPT solutions
are enough to compute at one-loop order in RPT
~\cite{Rampf12A, Rampf12B}.The contribution of the one-loop correction
is closed by 4LPT. 
 
Given the effects of higher-order LPT solutions,
RPT offers a more effective prediction method
for future deep-space surveys.

\subsection{Initial conditions for cosmological $N$-body simulations}
\label{subsec:init}

Although Lagrangian nonlinear perturbations describe quasi-nonlinear
evolution well, they cannot describe strongly nonlinear evolution.
After the formation of caustics, the hydrodynamical description no longer
has physical meaning.
For strongly nonlinear evolution, cosmological $N$-body simulations
have been used. With improvements to algorithms for the computation of
gravitational interactions and advancements in computer technology,
 large, detailed simulations can be carried out.

Although the growth of the density fluctuations starts in the recombination
era ($z \simeq 1000$), $N$-body simulations are begun from a later time,
because of the difficulty in performing the simulation with small fluctuations.
In the standard scheme, the small fluctuations evolve following the
perturbative approach until they reach the quasi-nonlinear regime
($z \simeq 30$)~\cite{COSMICS}.

In the perturbative approach, ZA has long been used.
In ZA, although the perturbation remains in the linear regime,
quasi-nonlinear density fluctuations can be described.
However, ZA does not reproduce the higher-order statistics,
because the acceleration is always parallel to the
peculiar velocity. In the quasilinear regime, effects that cannot be expressed
using only linear perturbations appear.
If we
set up the initial condition for use with a cosmological
$N$-body simulation at the quasi-nonlinear stage,
ZA seems insufficient for describing
the initial condition.

To solve this problem,
Crocce, Pueblas, and Scoccimarro~\cite{Crocce06B}
proposed an improvement in which different initial
conditions are adopted. Essentially, the initial conditions
are based on approximations that are valid up to
second-order LPT (2LPT) and
reproduce the exact value of the skewness
in the weakly nonlinear region.
They set up the initial conditions for ZA and 2LPT
at $z=49$. The simulations were run using the Gadget2 code
\cite{Gadget2} and contained $N=512^3$ particles.
The cosmological parameters used were
$\Omega_m=0.27,$ $\Omega_{\Lambda}=0.73,$ $\Omega_b=0.046,$
$h=0.72,$ and $\sigma_8(z=0) =0.9$. The box size was
$L_{{\rm box}}=512, 1024 h^{-1}$ [Mpc].

Crocce, Pueblas, and Scoccimarro showed the evolution of higher-order cumulants
such as skewness and kurtosis. First, they computed these
quantities for the reference runs (2LPT, $z_i=49$). Next,
they compared the quantities for reference runs with those for
several simulations performed using other initial conditions. The
difference between the skewness and the kurtosis values for
the reference runs and the simulations using ZA is
several percent for $2 < R < 60 h^{-1}$ [Mpc]
at $z=0, 1, 2, 3$.
Simulations using ZA underestimated the power spectrum
in the nonlinear regime by approximately 2\%, 4\%, and $8 \%$ at
$z = 0,$ 1, and $3$,
respectively.

Thereafter, the effect of 3LPT was verified by Tatekawa and Mizuno.
The conditions they considered differed from those in the former study.
The simulations were run using the $P^3M$ code~\cite{Bert1991}
and contained $N=128^3$ particles.
The cosmological parameters were taken from the
WMAP three-year results~\cite{WMAP-3yr}.
They considered only the longitudinal mode in 3LPT.
Their initial condition was given by $z \simeq 23, 33, 84$.
In their results, the difference of both the skewness and the kurtosis
appeared between the simulations using ZA and 2LPT.
Even if the initial condition is given at $z \simeq 84$, the
difference appeared. The result is consistent with that
by Crocce, Pueblas, and Scoccimarro.
On the other hand, 
if the initial condition was given by $z \simeq 33$,
the difference between both the skewness and the kurtosis
for the simulations using 2LPT and 3LPT is less than
$2 \%$ for $R=2 h^{-1}$ [Mpc] at late times.
According to their results,
2LPT is sufficiently accurate when several percentage points of precision
are needed for the higher-order cumulants.
In other words, the initial conditions given by 2LPT 
have the enough accuracy in simulations with precision of 
one percent.

In the future, when higher precision is needed, we should
select the initial conditions for $N$-body simulations using 3LPT
or 4LPT. The transverse mode of 3LPT was ignored in
previous analyses. Before estimating the effect of 4LPT,
it is necessary to estimate the effect of the transverse
mode in 3LPT.

\begin{figure}
     \centerline{\includegraphics[width=11cm]
                                     {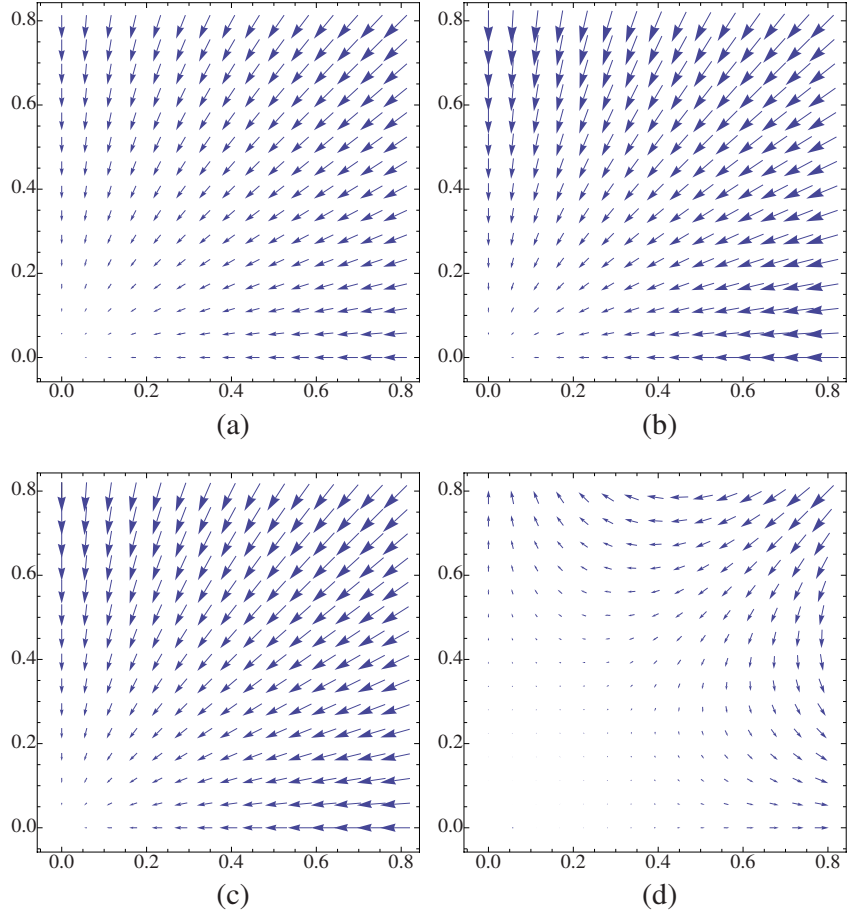}}
     \caption{The displacement generated by Eq.~(\ref{eqn:simple-ex}). In this case,
 $\psi_{,ii}^{(3b)}$ disappears.
 Here we plot in the range of $q_1, q_2 \in [ 0 , \pi/4]$.
 In these figures, the vectors are
 amplified. (a) 1LPT; (b) 2LPT; (c) 3LPT longitudinal mode; (d) 3LPT transverse mode.}
     \label{fig:Vector}
\end{figure}

Here we suppose simple perturbation
in first-order LPT (1LPT):
\begin{equation} \label{eqn:simple-ex}
\psi_1(\bm{q}) = \varepsilon \left \{ \cos q_1 + \cos q_2 \right \} .
\end{equation}
This perturbation gives the concentration of the matter.
The displacement vector is shown in Fig.~\ref{fig:Vector}.
The longitudinal
mode in 2LPT and 3LPT promotes the concentration. The direction of
the displacement in both 2LPT and the longitudinal mode in 3LPT  is parallel
to that in 1LPT. However, the displacement of the transverse
mode in 3LPT  differs from that of the others. This difference in the direction
of the displacement would especially affect the statistical quantities in
redshift space.

Here, we roughly estimate the effect of 4LPT.
The difference between the higher-order cumulants in the cases of
ZA and 2LPT was approximately $10 \%$. In contrast, the difference between the higher-order cumulants in the cases of 2LPT and 3LPT
was approximately $1 \%$. When the order of
the perturbation is raised, the difference is $1/10$.
This tendency suggests that when the desired precision is less than
$1 \%$, we should select the initial condition using 4LPT.

\section{Summary}
\label{sec:summary}

We have derived Lagrangian fourth-order perturbative equations for a
cosmological fluid. The equations consist of six scalar equations and four
vector equations. The spatial components of the perturbation are
described by Laplace's equation.

In this paper, although we assumed that the linear perturbation only had a
longitudinal mode, a transverse mode appears in the third-order perturbation;
we consider the effect of this transverse mode.
The density fluctuation is described by Eq.~(\ref{eqn:exact-delta}).
In 3LPT, even if the transverse mode is present in the third-order perturbation,
the mode does not affect the density fluctuation, because the
mode disappears in Eq.~(\ref{eqn:J-expansion}); this mode affects
only the displacement. The simple case was shown in Sec.~\ref{subsec:init}.

In 4LPT,
the doublet term consists of ZA, and the third-order transverse mode
appears. When the order of the approximation is raised to fourth order,
it is necessary to consider the effect of the transverse mode
on not only the displacement but also the density fluctuation.

The appearance of the transverse mode does not imply the occurrence of
vorticity. In LPT, the vorticity is described
exactly~\cite{Buchert1992,Barrow1993,Sasaki1998}:
\begin{equation}
\bm{\omega} = \frac{1}{a^2} \frac{ \left ( \bm{\omega} (\bm{q}, t_0) \cdot
 \nabla_q \right ) \bm{x} }{J} .
\end{equation}
$\omega (\bm{q}, t_0)$ refers to the initial vorticity. Therefore,
if  vorticity does not exist in the initial condition,
it never appears. This result is consistent with Kelvin's circulation theorem.

Here, we note a problem in higher-order perturbation.
If the primordial density fluctuation is negative,
the fluctuation $\delta$ approaches $-1$.
In Eulerian perturbation, the density fluctuation exceeds $-1$, i.e.,
negative mass density appears.
However, in LPT, $\delta$ decreases gently
and converges to $-1$.
However, the first-order perturbation remains the best approximation
for describing the late-time evolution of voids. In particular, if we stop
expansion when an even order (second, fourth, sixth, $\dots$) is
reached, the perturbative solution describes the contraction
of a void~\cite{Tatekawa2007,Munshi1994,Sahni1996}. Therefore, when
significantly advancing the time evolution, we should apply
the fourth-order perturbation.

Resummation theory significantly improves the precision of the perturbation.
Even if we consider fourth-order Lagrangian perturbation,
we could still avoid the above problem. In particular, the Lagrangian perturbation
easily computes physical quantities  not only in real space but also in redshift space. In the future, the resummation of the Lagrangian
perturbation up to the higher-order will contribute to the prediction methods used
in precision cosmology.

Rampf and Buchert~\cite{Rampf12A} have been derived 4LPT equations
independently.  Although their peculiar-system approach corresponds
to our approach,
their 4LPT equations are different from our equations.
The origin of the difference are treatment of temporal components and spatial
derivative. For temporal components, they introduced conformal time
for the equations.
Then they considered both growing and decaying modes.
For the spatial derivative,
they considered inverse matrix for the Jacobian matrix
between Eulerian coordinate and Lagrangian coordinate. 
At present, especially in the spatial components,
both equations are inconsistent. 
The meaning of the difference in the spatial terms
and its effects should be confirmed in near future.

For more higher-order correction, 4LPT would not be enough.
Enough higher-order correction to resummation, fifth-order LPT (5LPT) has been achieved. 
Based on the calculation of 4LPT, 5LPT equations will likely consist of at least 19
scalar and 18 vector equations with our approach. Although the derivation
of 5LPT equations appears to be quite a challenging problem, useful predictions for
future deep-space observation might be possible; therefore, it is worth wrestling
with this problem.

\section*{Acknowledgments}
We thank Takahiko Matsubara for carefully reading
this manuscript and providing advice
regarding the recent
developments in resummation theory.
We also thank Shuntaro Mizuno for useful discussions.

\appendix
\section{Determinant of displacement}\label{sec:det}
We consider rewriting the determinant of the Lagrangian displacement matrix
with triplet terms.

First, we define $\mathcal{D}$ as follows:
\begin{eqnarray}
\mathcal{D} \left (A, B, C \right )
 & \equiv &
 A_{,11} B_{,22} C_{,33} + A_{,12} B_{,23} C_{,31} + A_{,13} B_{,21} C_{,32}
 \nonumber \\
 && - A_{,11} B_{,23} C_{,32} + A_{,22} B_{,31} C_{,13} + A_{,33} B_{,12} C_{,21}
 \nonumber \\
 && + A_{,33} B_{,11} C_{,22} + A_{,31} B_{,12} C_{,23} + A_{,32} B_{,13} C_{,21}
 \nonumber \\
 && - A_{,32} B_{,11} C_{,23} + A_{,13} B_{,22} C_{,31} + A_{,21} B_{,33} C_{,12}
 \nonumber \\
 && + A_{,22} B_{,33} C_{,11} + A_{,23} B_{,31} C_{,12} + A_{,21} B_{,32} C_{,13}
 \nonumber \\
 && - A_{,23} B_{,32} C_{,11} + A_{,31} B_{,13} C_{,22} + A_{,12} B_{,21} C_{,33}
 . \nonumber \\
&=& A_{,ii} B_{,jj} C_{,kk} + A_{,ij} B_{,jk} C_{,ki}
 + A_{,ij} B_{,ki} C_{,jk}
 \nonumber \\
 && - \left ( A_{,ii} B_{,jk} C_{,kj}
 + A_{,kj} B_{,ii} C_{,jk} + A_{,jk} B_{,kj} C_{,ii} \right ) ,
\end{eqnarray}
\begin{eqnarray}
&& \mathcal{D} \left (A, B, C \right )
 = \mathcal{D} \left (B, C, A \right ) =
\mathcal{D} \left (C, A, B \right ) \nonumber \\
&=& \mathcal{D} \left (A, C, B \right ) =
\mathcal{D} \left (B, A, C \right ) = \mathcal{D} \left (C, B, A \right ) ,
\end{eqnarray}
\begin{equation}
\mathcal{D} \left (A, A, A \right ) = 6 \det \left ( A_{,ij} \right ) ,
\end{equation}
where $A$, $B,$ and $C$ are scalar functions of $\bm{q}$.
In this paper, we consider fourth-order perturbation. Therefore, we
must consider a determinant that consists of first- and second-order
perturbations:
\begin{eqnarray}
\det \left (S_{,ij}^{(1)} + S_{,ij}^{(2)} \right )
&=& \frac{1}{6} \mathcal{D} \left ( S^{(1)}+S^{(2)}, S^{(1)}+S^{(2)},
 S^{(1)}+S^{(2)} \right ) \nonumber \\
&=& \frac{1}{6} \left [ \mathcal{D} \left ( S^{(1)}, S^{(1)}, S^{(1)} \right )
 + \mathcal{D} \left ( S^{(2)}, S^{(1)}, S^{(1)} \right )
 \right . \nonumber \\
&& ~~~~ \left .
 + \mathcal{D} \left ( S^{(1)}, S^{(2)}, S^{(1)} \right )
 + \mathcal{D} \left ( S^{(1)}, S^{(1)}, S^{(2)} \right ) \right ]
\nonumber \\
&& + O(\varepsilon^5) \nonumber \\
&=& \frac{1}{6} \mathcal{D} \left ( S^{(1)}, S^{(1)}, S^{(1)} \right )
 + \frac{1}{2} \mathcal{D} \left ( S^{(1)}, S^{(1)}, S^{(2)} \right )
  + O(\varepsilon^5) .
\end{eqnarray}
The determinant is rewritten in triplet terms as
\begin{eqnarray}
\det \left ( S_{,ij}^{(1)} \right )
 &=& \frac{1}{6} \mathcal{D} \left (S^{(1)}, S^{(1)}, S^{(1)} \right )
 \nonumber \\
 &=& \frac{1}{6} S_{,ii}^{(1)} S_{,jj}^{(1)} S_{,kk}^{(1)}
 - \frac{1}{2} S_{,ii}^{(1)} S_{,jk}^{(1)} S_{,jk}^{(1)}
 + \frac{1}{3} S_{,ij}^{(1)} S_{,jk}^{(1)} S_{,ki}^{(1)} , \\
\frac{1}{2} \mathcal{D} \left (S^{(1)}, S^{(1)}, S^{(2)} \right )
 &=& \frac{1}{2} \left [ \left \{ S_{,ii}^{(1)} S_{,jj}^{(1)}
 - S_{,ij}^{(1)} S_{,ij}^{(1)} \right \} S_{,kk}^{(2)}
 \right . \nonumber \\
 && ~~~~ \left . + 2 \left \{ S_{,ij}^{(1)} S_{,jk}^{(1)} S_{,ki}^{(2)}
  - S_{,ii}^{(1)} S_{,jk}^{(1)} S_{,jk}^{(2)} \right \}
 \right ] .
\end{eqnarray}

\section{Derivation of fourth-order perturbative equations}\label{sec:dev-4th}

In this section, we provide a detailed description of the derivation of fourth-order perturbative equations. First, we present a derivation of the longitudinal mode.
The left-hand-side terms of the evolution equation (\ref{eqn:L-longi-eqn}) for
the fourth-order longitudinal mode are given as follows:
\begin{eqnarray}
&& \ddot{S}_{,ii}^{(4)} + 2 \frac{\dot{a}}{a} \dot{S}_{,ii}^{(4)}
 - s_{i,j}^{(3)} \left ( \ddot{S}_{,ij}^{(1)}
 + 2 \frac{\dot{a}}{a} \dot{S}_{,ij}^{(1)} \right )
 - S_{,ij}^{(2)} \left ( \ddot{S}_{,ij}^{(2)}
 + 2 \frac{\dot{a}}{a} \dot{S}_{,ij}^{(2)} \right )
 \nonumber \\
&-& S_{,ij}^{(1)} \left ( \ddot{s}_{i,j}^{(3)}
 + 2 \frac{\dot{a}}{a} \dot{s}_{i,j}^{(3)} \right )
 + S_{,ij}^{(1)} S_{,jk}^{(1)}
 \left ( \ddot{S}_{,ki}^{(2)}
 + 2 \frac{\dot{a}}{a} \dot{S}_{,ki}^{(2)} \right )
 \nonumber \\
&+& S_{,ij}^{(1)} S_{,jk}^{(2)}
 \left ( \ddot{S}_{,ki}^{(1)}
 + 2 \frac{\dot{a}}{a} \dot{S}_{,ki}^{(1)} \right )
 + S_{,ij}^{(2)} S_{,jk}^{(1)}
 \left ( \ddot{S}_{,ki}^{(1)}
 + 2 \frac{\dot{a}}{a} \dot{S}_{,ki}^{(1)} \right )
 \nonumber \\
&-& S_{,ij}^{(1)} S_{,jk}^{(1)} S_{,kl}^{(1)}
 \left ( \ddot{S}_{,li}^{(1)}
 + 2 \frac{\dot{a}}{a} \dot{S}_{,li}^{(1)} \right ) .
\end{eqnarray}
The right-hand-side terms of the evolution equation (\ref{eqn:L-longi-eqn}) for
the fourth-order longitudinal mode are given as follows:
\begin{eqnarray}
-4 \pi G \rho_b & \times &
 \left [ -S_{,ii}^{(4)} + 2 S_{,ii}^{(1)} S_{,jj}^{(3)}
 + S_{,ii}^{(2)} S_{,jj}^{(2)}
 - \frac{1}{2} \left \{ S_{,ii}^{(2)} S_{,jj}^{(2)}
  - S_{,ij}^{(2)} S_{,ij}^{(2)} \right \}
 \right . \nonumber \\
 && ~~ - \left \{ S_{,ii}^{(1)} S_{,jj}^{(3)}
  - S_{,ij}^{(1)} s_{i,j}^{(3)} \right \}
  - \frac{1}{2} \mathcal{D} \left (S^{(1)}, S^{(1)}, S^{(2)} \right )
 \nonumber \\
 && ~~ + \frac{1}{4}
 \left \{ S_{,ii}^{(1)} S_{,jj}^{(1)} - S_{,ij}^{(1)} S_{,ij}^{(1)}
 \right \}
 \left \{ S_{,kk}^{(1)} S_{,ll}^{(1)} - S_{,kl}^{(1)} S_{,kl}^{(1)}
 \right \}
 \nonumber \\
 && ~~ + 2 S_{,ii}^{(1)} \left ( S_{,jj}^{(1)} S_{,kk}^{(2)}
  - S_{,jk}^{(1)} S_{,jk}^{(2)} \right )
  + S_{,ii}^{(2)} \left ( S_{,jj}^{(1)} S_{,kk}^{(1)}
  - S_{,jk}^{(1)} S_{,jk}^{(1)} \right )
 \nonumber \\
 && ~~ + 2 S_{,ii}^{(1)} \det \left (S_{,jk}^{(1)} \right )
  - 3 S_{,ii}^{(1)} S_{,jj}^{(1)} S_{,kk}^{(2)}
 \nonumber \\
 && ~~ \left . - \frac{3}{2} S_{,ii}^{(1)} S_{,jj}^{(1)}
 \left \{ S_{,kk}^{(1)} S_{,ll}^{(1)} - S_{,kl}^{(1)} S_{,kl}^{(1)}
 \right \}
  + S_{,ii}^{(1)} S_{,jj}^{(1)} S_{,kk}^{(1)} S_{,ll}^{(1)}
 \right ] .
\end{eqnarray}
The definition of $\mathcal{D}$ is given in Appendix \ref{sec:det}.
We rewrite the homogeneous terms in the fourth-order perturbative
equation using first-, second-, and third-order equations
(\ref{eqn:1st-L-eq})--(\ref{eqn:1st-L-t-eq}),
(\ref{eqn:2nd-L-ts})--(\ref{eqn:2nd-L-spa-eq}),
(\ref{eqn:3rd-L-t-eqA})--(\ref{eqn:3rd-L-spa-eqB}), and
(\ref{eqn:3rd-T-ts}), respectively.
First, we examine the doublet terms in the first- and third-order
perturbations:
\begin{eqnarray}
&& s_{i,j}^{(3)} \left ( \ddot{S}_{,ij}^{(1)}
 + 2 \frac{\dot{a}}{a} \dot{S}_{,ij}^{(1)} \right )
 + S_{,ij}^{(1)} \left ( \ddot{s}_{i,j}^{(3)}
 + 2 \frac{\dot{a}}{a} \dot{s}_{i,j}^{(3)} \right ) \nonumber \\
&& - 4 \pi G \rho_b \left [ 2 S_{,ii}^{(1)} S_{,jj}^{(3)}
  - \left \{ S_{,ii}^{(1)} S_{,jj}^{(3)}
  - S_{,ij}^{(1)} s_{i,j}^{(3)} \right \} \right ] \nonumber \\
&=& 4 \pi G \rho_b \times \left [ g_1 g_{3a} \psi_{,ij}^{(1)}
 \psi_{,ij}^{(3a)}
 + g_1 g_{3b} \psi_{,ij}^{(1)} \psi_{,ij}^{(3b)}
 + g_1 g_{3T} \psi_{,ij}^{(1)} \zeta_{i,j}^{(3)} \right . \nonumber \\
&& ~~~~~~
 + g_1 ( g_{3a} - 2g_1 g_2 + 2 g_1^3 ) \psi_{,ij}^{(1)} \psi_{,ij}^{(3a)}
 + g_1 ( g_{3b} - 2 g_1^3 ) \psi_{,ij}^{(1)} \psi_{,ij}^{(3b)}
 + g_1^4 \psi_{,ij}^{(1)} \zeta_{i,j}^{(3)} \nonumber \\
&& ~~~~~~
 - 2 g_1 g_{3a} \psi_{,ii}^{(1)} \psi_{,jj}^{(3a)}
 - 2 g_1 g_{3b} \psi_{,ii}^{(1)} \psi_{,jj}^{(3b)}
\nonumber \\
&& ~~~~~~
 + g_1 g_{3a} \left ( \psi_{,ii}^{(1)} \psi_{,jj}^{(3a)} - \psi_{,ij}^{(1)}
 \psi_{,ij}^{(3a)} \right )
 + g_1 g_{3b} \left ( \psi_{,ii}^{(1)} \psi_{,jj}^{(3b)} - \psi_{,ij}^{(1)}
 \psi_{,ij}^{(3b)} \right )
\nonumber \\
&& ~~~~~~ \left .
 - g_1 g_{3T} \psi_{,ij}^{(1)} \zeta_{i,j}^{(3)} \right ] \nonumber \\
&=& - 8 \pi G \rho_b g_1 ( g_{3a} - 2g_1 g_2 + 2 g_1^3 ) \cdot
 \frac{1}{2} \left ( \psi_{,ii}^{(1)} \psi_{,jj}^{(3a)} - \psi_{,ij}^{(1)}
 \psi_{,ij}^{(3a)} \right ) \nonumber \\
&& - 8 \pi G \rho_b g_1 ( g_{3b} - 2 g_1^3 ) \cdot
 \frac{1}{2} \left ( \psi_{,ii}^{(1)} \psi_{,jj}^{(3b)} - \psi_{,ij}^{(1)}
 \psi_{,ij}^{(3b)} \right )
 + 4 \pi G \rho_b g_1^4 \psi_{,ij}^{(1)} \zeta_{i,j}^{(3)} \nonumber \\
&& + 4 \pi G \rho_b g_1 (-2 g_1 g_2 + 2 g_1^3)
 \psi_{,ii}^{(1)} \psi_{,jj}^{(3a)}
 - 8 \pi G \rho_b g_1^4 \psi_{,ii}^{(1)} \psi_{,jj}^{(3b)} \nonumber \\
&=& - 8 \pi G \rho_b g_1 ( g_{3a} - 2g_1 g_2 + 2 g_1^3 ) \cdot
 \frac{1}{2} \left ( \psi_{,ii}^{(1)} \psi_{,jj}^{(3a)} - \psi_{,ij}^{(1)}
 \psi_{,ij}^{(3a)} \right ) \nonumber \\
&& - 8 \pi G \rho_b g_1 ( g_{3b} - 2 g_1^3 ) \cdot
 \frac{1}{2} \left ( \psi_{,ii}^{(1)} \psi_{,jj}^{(3b)} - \psi_{,ij}^{(1)}
 \psi_{,ij}^{(3b)} \right )
 + 4 \pi G \rho_b g_1^4 \psi_{,ij}^{(1)} \zeta_{i,j}^{(3)} \nonumber \\
&& + 4 \pi G \rho_b g_1^2 (-2 g_2 + 2 g_1^2 ) \cdot \frac{1}{2}
 \psi_{,ii}^{(1)} \left \{
 \psi_{,jj}^{(1)} \psi_{,kk}^{(2)} - \psi_{,jk}^{(1)}
 \psi_{,jk}^{(2)} \right \}
 \nonumber \\
&& - 8 \pi G \rho_b g_1^4 \psi_{,ii}^{(1)}
\det \left (\psi_{,jk}^{(1)} \right ) . \label{eqn:4th-L-source-1-3}
\end{eqnarray}
The upper two lines on the right-hand side of Eq.~(\ref{eqn:4th-L-source-1-3})
correspond to the source terms of
Eqs.~(\ref{eqn:4th-L-g4a})--(\ref{eqn:4th-L-g4c}) and
Eqs.~(\ref{eqn:4th-L-psi4a})--(\ref{eqn:4th-L-psi4c}).
The other lines (remainder terms) in Eq.~(\ref{eqn:4th-L-source-1-3})
will be considered later.
Next, we examine the doublet terms of the second-order perturbations:
\begin{eqnarray}
&& S_{,ij}^{(2)} \left ( \ddot{S}_{,ij}^{(2)}
 + 2 \frac{\dot{a}}{a} \dot{S}_{,ij}^{(2)} \right )
 - 4 \pi G \rho_b \left [ S_{,ii}^{(2)} S_{,jj}^{(2)}
 - \frac{1}{2} \left \{ S_{,ii}^{(2)} S_{,jj}^{(2)}
  - S_{,ij}^{(2)} S_{,ij}^{(2)} \right \} \right ] \nonumber \\
&=& 4 \pi G \rho_b \times \left [ g_2 (g_2 - g_1^2)
 \psi_{,ij}^{(2)} \psi_{,ij}^{(2)}
 - g_2^2 \psi_{,ii}^{(2)} \psi_{,jj}^{(2)}
 + \frac{1}{2} g_2^2 \left \{ \psi_{,ii}^{(2)} \psi_{,jj}^{(2)}
 - \psi_{,ij}^{(2)} \psi_{,ij}^{(2)} \right \}
 \right ] \nonumber \\
&=& -4 \pi G \rho_b g_2 (g_2 - 2 g_1^2) \cdot \frac{1}{2}
 \left \{ \psi_{,ii}^{(2)} \psi_{,jj}^{(2)}
 - \psi_{,ij}^{(2)} \psi_{,ij}^{(2)} \right \}
 -4 \pi G \rho_b g_1^2 g_2 \cdot \frac{1}{2}
 \psi_{,ii}^{(2)} \psi_{,jj}^{(2)} \nonumber \\
&=& -4 \pi G \rho_b g_2 (g_2 - 2 g_1^2) \cdot \frac{1}{2}
 \left \{ \psi_{,ii}^{(2)} \psi_{,jj}^{(2)}
 - \psi_{,ij}^{(2)} \psi_{,ij}^{(2)} \right \}
 \nonumber \\
&& -4 \pi G \rho_b g_1^2 g_2 \cdot \frac{1}{4}
 \psi_{,ii}^{(2)} \left \{ \psi_{,jj}^{(1)} \psi_{,kk}^{(1)}
 - \psi_{,jk}^{(1)} \psi_{,jk}^{(1)}
 \right \}
. \label{eqn:4th-L-source-2-2}
\end{eqnarray}
The first line on the right-hand side of Eq.~(\ref{eqn:4th-L-source-2-2})
corresponds to the source terms of
Eqs.~(\ref{eqn:4th-L-g4d}) and (\ref{eqn:4th-L-psi4d}).
The last terms (remainder terms) in Eq.~(\ref{eqn:4th-L-source-2-2})
will be considered later.
We notice the triplet terms in the first- and second-order perturbations:
\begin{eqnarray}
&& - S_{,ij}^{(1)} S_{,jk}^{(1)} \left ( \ddot{S}_{,ki}^{(2)}
 + 2 \frac{\dot{a}}{a} \dot{S}_{,ki}^{(2)} \right )
 - S_{,ij}^{(1)} S_{,jk}^{(2)}
 \left ( \ddot{S}_{,ki}^{(1)}
 + 2 \frac{\dot{a}}{a} \dot{S}_{,ki}^{(1)} \right )
 \nonumber \\
&& - S_{,ij}^{(2)} S_{,jk}^{(1)}
 \left ( \ddot{S}_{,ki}^{(1)}
 + 2 \frac{\dot{a}}{a} \dot{S}_{,ki}^{(1)} \right )
 \nonumber \\
&& + 4 \pi G \rho_b \times \left [
 \mathcal{D} \left (S^{(1)}, S^{(1)}, S^{(2)} \right )
 - 2 S_{,ii}^{(1)} \left ( S_{,jj}^{(1)} S_{,kk}^{(2)}
  - S_{,jk}^{(1)} S_{,jk}^{(2)} \right )
 \right . \nonumber \\
&& ~~~~~~~~~~~~~~ \left .
 - S_{,ii}^{(2)} \left ( S_{,jj}^{(1)} S_{,kk}^{(1)}
  - S_{,jk}^{(1)} S_{,jk}^{(1)} \right )
 + 3 S_{,ii}^{(1)} S_{,jj}^{(1)} S_{,kk}^{(2)}
  \right ] \nonumber \\
&=& 4 \pi G \rho_b \times \left [ -g_1^2 (3 g_2 - g_1^2)
  \psi_{,ij}^{(1)} \psi_{,jk}^{(1)} \psi_{,ki}^{(2)}
  + g_1^2 g_2 \mathcal{D} \left (S^{(1)}, S^{(1)}, S^{(2)} \right )
 \right .  \nonumber \\
&& ~~~~~~~~~~~~~~ \left .
 + 2 g_1^2 g_2 \psi_{,ii}^{(1)} \psi_{,jk}^{(1)} \psi_{,jk}^{(2)}
 + g_1^2 g_2 \psi_{,ii}^{(2)} \psi_{,jk}^{(1)} \psi_{,jk}^{(1)}
 \right ] \nonumber \\
 &=& 4 \pi G \rho_b \times \left [ -g_1^2 (2 g_2 - g_1^2)
  \psi_{,ij}^{(1)} \psi_{,jk}^{(1)} \psi_{,ki}^{(2)}
  + g_1^2 g_2 \psi_{,ii}^{(1)} \psi_{,jk}^{(1)} \psi_{,jk}^{(2)}
 \right .  \nonumber \\
&& ~~~~~~~~~~~~~~ \left .
  + \frac{1}{2} g_1^2 g_2
  \psi_{ii}^{(2)} \psi_{,jk}^{(1)} \psi_{,jk}^{(1)}
  + \frac{1}{2} g_1^2 g_2 \psi_{,ii}^{(2)} \psi_{,jj}^{(1)} \psi_{,kk}^{(1)}
 \right ] .
\end{eqnarray}
Here, we add the remainder terms in Eqs.~(\ref{eqn:4th-L-source-1-3})
and (\ref{eqn:4th-L-source-2-2}). The triplet terms become
\begin{eqnarray}
&& 4 \pi G \rho_b g_1^2 \times \left [ (-g_2 + g_1^2)
 \psi_{,ii}^{(1)} \left \{
 \psi_{,jj}^{(1)} \psi_{,kk}^{(2)} - \psi_{,jk}^{(1)}
 \psi_{,jk}^{(2)} \right \} \right . \nonumber \\
&& ~~~~~~~~~~~~ - \frac{1}{4} g_2
 \psi_{,ii}^{(2)} \left \{ \psi_{,jj}^{(1)} \psi_{,kk}^{(1)}
 - \psi_{,jk}^{(1)} \psi_{,jk}^{(1)}
 \right \} \nonumber \\
&& ~~~~~~~~~~~~ - (2 g_2-g_1^2)
  \psi_{,ij}^{(1)} \psi_{,jk}^{(1)} \psi_{,ki}^{(2)}
  + g_2 \psi_{,ii}^{(1)} \psi_{,jk}^{(1)} \psi_{,jk}^{(2)}
 \nonumber \\
&& ~~~~~~~~~~~~ \left .
 + \frac{1}{2} g_2 \psi_{,ii}^{(2)} \psi_{,jk}^{(1)} \psi_{,jk}^{(1)}
  + \frac{1}{2} g_2 \psi_{,ii}^{(2)} \psi_{,jj}^{(1)} \psi_{,kk}^{(1)}
 \right ] \nonumber \\
&=& 4 \pi G \rho_b g_1^2 (2 g_2 - g_1^2)
 \left [ \psi_{,ii}^{(1)} \psi_{,jk}^{(1)} \psi_{,jk}^{(2)}
 - \psi_{,ij}^{(1)} \psi_{,jk}^{(1)} \psi_{,ki}^{(2)}
 - \frac{3}{8} \psi_{,ii}^{(2)}
  \left \{ \psi_{,jj}^{(1)} \psi_{,kk}^{(1)}
 - \psi_{,jk}^{(1)} \psi_{,jk}^{(1)} \right \} \right ] \nonumber \\
&&  + 4 \pi G \rho_b g_1^4 \psi_{,ii}^{(2)}
 \left \{ \frac{5}{8} \psi_{,jj}^{(1)} \psi_{,kk}^{(1)}
 + \frac{3}{8} \psi_{,jk}^{(1)} \psi_{,jk}^{(1)} \right \}
 . \label{eqn:4th-L-source-1-1-2}
\end{eqnarray}
The first line on the right-hand side of Eq.~(\ref{eqn:4th-L-source-1-1-2})
corresponds to the source terms of
Eqs.~(\ref{eqn:4th-L-g4e}) and (\ref{eqn:4th-L-psi4e}).
The last terms (remainder terms) in Eq.~(\ref{eqn:4th-L-source-1-1-2}) will be considered.
Finally, we consider the quartet terms in the first-order perturbations:
\begin{eqnarray}
&&  S_{,ij}^{(1)} S_{,jk}^{(1)} S_{,kl}^{(1)}
 \left ( \ddot{S}_{,li}^{(1)}
 + 2 \frac{\dot{a}}{a} \dot{S}_{,li}^{(1)} \right ) \nonumber \\
&& + 4 \pi G \rho_b \times \left [
 - \frac{1}{4}
 \left \{ S_{,ii}^{(1)} S_{,jj}^{(1)} - S_{,ij}^{(1)} S_{,ij}^{(1)}
 \right \}
 \left \{ S_{,kk}^{(1)} S_{,ll}^{(1)} - S_{,kl}^{(1)} S_{,kl}^{(1)}
 \right \}
 - 2 S_{,ii}^{(1)} \det \left (S_{,jk}^{(1)} \right )
 \right . \nonumber \\
&& ~~~~~~~~~~~~ \left .
 + \frac{3}{2} S_{,ii}^{(1)} S_{,jj}^{(1)}
 \left \{ S_{,kk}^{(1)} S_{,ll}^{(1)} - S_{,kl}^{(1)} S_{,kl}^{(1)}
 \right \}
  + S_{,ii}^{(1)} S_{,jj}^{(1)} S_{,kk}^{(1)} S_{,ll}^{(1)}
 \right ] \nonumber \\
&=& 4 \pi G \rho_b g_1^4 \times \left [
 \psi_{,ij}^{(1)} \psi_{,jk}^{(1)} \psi_{,kl}^{(1)} \psi_{,li}^{(1)}
 - \frac{1}{4} \left \{ \psi_{,ii}^{(1)} \psi_{,jj}^{(1)}
   - \psi_{,ij}^{(1)} \psi_{,ij}^{(1)} \right \}
  \left \{ \psi_{,kk}^{(1)} \psi_{,ll}^{(1)}
   - \psi_{,kl}^{(1)} \psi_{,kl}^{(1)} \right \} \right . \nonumber \\
&& ~~~~~~~~~~~~ - 2 \psi_{,ii}^{(1)}
 \det \left ( \psi_{,jk}^{(1)} \right )
+ \frac{3}{2} \psi_{,ii}^{(1)} \psi_{,jj}^{(1)}
  \left \{ \psi_{,kk}^{(1)} \psi_{,ll}^{(1)}
   - \psi_{,kl}^{(1)} \psi_{,kl}^{(1)} \right \} \nonumber \\
&& ~~~~~~~~~~~~ \left .
 + \psi_{,ii}^{(1)} \psi_{,jj}^{(1)} \psi_{,kk}^{(1)} \psi_{,ll}^{(1)}
  \right ] .
\end{eqnarray}
Here, we add the remainder terms of
Eq.~(\ref{eqn:4th-L-source-1-1-2}).
The quartet terms are rewritten as follows:
\begin{eqnarray}
&& 4 \pi G \rho_b g_1^4 \times \left [
 \psi_{,ij}^{(1)} \psi_{,jk}^{(1)} \psi_{,kl}^{(1)} \psi_{,li}^{(1)}
 \right . \nonumber \\
&& ~~~~~~~~~~~~~~
 - \frac{1}{4} \left \{ \psi_{,ii}^{(1)} \psi_{,jj}^{(1)}
   - \psi_{,ij}^{(1)} \psi_{,ij}^{(1)} \right \}
  \left \{ \psi_{,kk}^{(1)} \psi_{,ll}^{(1)}
   - \psi_{,kl}^{(1)} \psi_{,kl}^{(1)} \right \} \nonumber \\
&& ~~~~~~~~~~~~~~ - 2 \psi_{,ii}^{(1)}
 \det \left ( \psi_{,jk}^{(1)} \right )
 + \frac{3}{2} \psi_{,ii}^{(1)} \psi_{,jj}^{(1)}
  \left \{ \psi_{,kk}^{(1)} \psi_{,ll}^{(1)}
   - \psi_{,kl}^{(1)} \psi_{,kl}^{(1)} \right \} \nonumber \\
&& ~~~~~~~~~~~~~~
 + \psi_{,ii}^{(1)} \psi_{,jj}^{(1)} \psi_{,kk}^{(1)} \psi_{,ll}^{(1)}
 - 2 \psi_{,ii}^{(1)} \det \left ( \psi_{,jk}^{(1)} \right )
\nonumber \\
&& ~~~~~~~~~~~~~~ + \frac{5}{16} \psi_{,ii}^{(1)} \psi_{,jj}^{(1)}
   \left \{ \psi_{,kk}^{(1)} \psi_{,ll}^{(1)}
   - \psi_{,kl}^{(1)} \psi_{,kl}^{(1)} \right \} \nonumber \\
&& ~~~~~~~~~~~~~~ \left .
 + \frac{3}{16} \psi_{,ij}^{(1)} \psi_{,ij}^{(1)}
  \left \{ \psi_{,kk}^{(1)} \psi_{,ll}^{(1)}
   - \psi_{,kl}^{(1)} \psi_{,kl}^{(1)} \right \} \right ]
  \nonumber \\
&=& 4 \pi G \rho_b g_1^4 \times \left [
 \psi_{,ij}^{(1)} \psi_{,jk}^{(1)} \psi_{,kl}^{(1)} \psi_{,li}^{(1)}
 - \frac{7}{16} \left \{ \psi_{,ii}^{(1)} \psi_{,jj}^{(1)}
   - \psi_{,ij}^{(1)} \psi_{,ij}^{(1)} \right \}
  \left \{ \psi_{,kk}^{(1)} \psi_{,ll}^{(1)}
   - \psi_{,kl}^{(1)} \psi_{,kl}^{(1)} \right \} \right . \nonumber \\
&& ~~~~~~~~~~~~~~ - 4 \psi_{,ii}^{(1)}
 \det \left ( \psi_{,jk}^{(1)} \right ) + 2 \psi_{,ii}^{(1)} \psi_{,jj}^{(1)}
  \left \{ \psi_{,kk}^{(1)} \psi_{,ll}^{(1)}
   - \psi_{,kl}^{(1)} \psi_{,kl}^{(1)} \right \} \nonumber \\
&& ~~~~~~~~~~~~~~ \left .
 + \psi_{,ii}^{(1)} \psi_{,jj}^{(1)} \psi_{,kk}^{(1)} \psi_{,ll}^{(1)}
  \right ] \nonumber \\
&=& 4 \pi G \rho_b g_1^4 \times \left [
 - \frac{7}{16} \left \{ \psi_{,ii}^{(1)} \psi_{,jj}^{(1)}
   - \psi_{,ij}^{(1)} \psi_{,ij}^{(1)} \right \}
  \left \{ \psi_{,kk}^{(1)} \psi_{,ll}^{(1)}
   - \psi_{,kl}^{(1)} \psi_{,kl}^{(1)} \right \} \right . \nonumber \\
&& ~~~~~~~~~~~~~~ \left . + \frac{7}{3} \psi_{,ii}^{(1)} \psi_{,jj}^{(1)}
  \psi_{,kk}^{(1)} \psi_{,ll}^{(1)} -
 \frac{4}{3} \psi_{,ii}^{(1)} \psi_{,jk}^{(1)} \psi_{,kl}^{(1)}
 \psi_{,li}^{(1)}
 + \psi_{,ij}^{(1)} \psi_{,jk}^{(1)} \psi_{,kl}^{(1)} \psi_{,li}^{(1)}
  \right ] . \label{eqn:4th-L-source-quar1}
\end{eqnarray}
The right-hand side of Eq.~(\ref{eqn:4th-L-source-quar1})
corresponds to the source terms of
Eqs.~(\ref{eqn:4th-L-g4f}) and (\ref{eqn:4th-L-psi4f}).

Next, we derive the equations that describe the transverse mode.
The evolution equation is derived from Eq.~(\ref{eqn:L-trans-eqn}) as
\begin{eqnarray}
&& \varepsilon_{ijk}
 \left (\ddot{s}_{k,j}^{T (4)} + 2 \frac{\dot{a}}{a}
 \dot{s}_{k,j}^{T (4)} \right )
- \varepsilon_{ijk}
 S_{,jl}^{(1)} \left ( \ddot{s}_{k,l}^{(3)}
+ 2 \frac{\dot{a}}{a} \dot{s}_{k,l}^{(3)} \right )
 \nonumber \\
&& - \varepsilon_{ijk}
 S_{,jl}^{(2)} \left ( \ddot{S}_{,kl}^{(2)}
+ 2 \frac{\dot{a}}{a} \dot{S}_{,kl}^{(2)} \right )
- \varepsilon_{ijk}
 s_{j,l}^{(3)} \left ( \ddot{S}_{,kl}^{(1)}
+ 2 \frac{\dot{a}}{a} \dot{S}_{,kl}^{(1)} \right ) \nonumber \\
&& + \varepsilon_{ijk}
 S_{,jl}^{(1)} S_{,lm}^{(1)} \left ( \ddot{S}_{,km}^{(2)}
+ 2 \frac{\dot{a}}{a} \dot{S}_{,km}^{(2)} \right )
 + \varepsilon_{ijk}
 S_{,jl}^{(1)} S_{,lm}^{(2)} \left ( \ddot{S}_{,km}^{(1)}
+ 2 \frac{\dot{a}}{a} \dot{S}_{,km}^{(1)} \right )
\nonumber \\
&& + \varepsilon_{ijk}
 S_{,jl}^{(2)} S_{,lm}^{(1)} \left ( \ddot{S}_{,km}^{(1)}
+ 2 \frac{\dot{a}}{a} \dot{S}_{,km}^{(1)} \right )
 - \varepsilon_{ijk}
 S_{,jl}^{(1)} S_{,lm}^{(1)} S_{,mn}^{(1)} \left ( \ddot{S}_{,kn}^{(1)}
+ 2 \frac{\dot{a}}{a} \dot{S}_{,kn}^{(1)} \right )
\nonumber \\
 &=& 0 . \label{eqn:4th-T}
\end{eqnarray}
Three terms on the left-hand side of Eq.~(\ref{eqn:4th-T}) disappear
owing to the symmetry of subscripts in the longitudinal mode, and so
\begin{eqnarray}
\varepsilon_{ijk} S_{,jl}^{(2)} \left ( \ddot{S}_{,kl}^{(2)}
+ 2 \frac{\dot{a}}{a} \dot{S}_{,kl}^{(2)} \right )
 &=& 4 \pi G \rho_b g_2^2 \varepsilon_{ijk}
     \psi_{,jl}^{(2)} \psi_{,kl}^{(2)} = 0 , \\
\varepsilon_{ijk} S_{,jl}^{(1)} S_{,lm}^{(2)} \left ( \ddot{S}_{,km}^{(1)}
+ 2 \frac{\dot{a}}{a} \dot{S}_{,km}^{(1)} \right )
 &=& 4 \pi G \rho_b g_1^2 g_2 \varepsilon_{ijk}
     \psi_{,jl}^{(1)} \psi_{,km}^{(1)} \psi_{,lm}^{(2)} = 0 , \\
\varepsilon_{ijk}
 S_{,jl}^{(1)} S_{,lm}^{(1)} S_{,mn}^{(1)} \left ( \ddot{S}_{,kn}^{(1)}
+ 2 \frac{\dot{a}}{a} \dot{S}_{,kn}^{(1)} \right )
 &=& 4 \pi G \rho_b g_1^4 \varepsilon_{ijk}
    \psi_{,jl}^{(1)} \psi_{,kn}^{(1)} \psi_{,lm}^{(1)} \psi_{,mn}^{(1)}
 =0 .
\end{eqnarray}
Equation~(\ref{eqn:4th-T}) is rewritten as follows:
\begin{eqnarray}
&& \varepsilon_{ijk}
 \left (\ddot{s}_{k,j}^{T (4)} + 2 \frac{\dot{a}}{a}
 \dot{s}_{k,j}^{T (4)} \right )
- \varepsilon_{ijk}
 S_{,jl}^{(1)} \left ( \ddot{s}_{k,l}^{(3)}
+ 2 \frac{\dot{a}}{a} \dot{s}_{k,l}^{(3)} \right )
 - \varepsilon_{ijk}
 s_{j,l}^{(3)} \left ( \ddot{S}_{,kl}^{(1)}
+ 2 \frac{\dot{a}}{a} \dot{S}_{,kl}^{(1)} \right ) \nonumber \\
&& + \varepsilon_{ijk}
 S_{,jl}^{(1)} S_{,lm}^{(1)} \left ( \ddot{S}_{,km}^{(2)}
+ 2 \frac{\dot{a}}{a} \dot{S}_{,km}^{(2)} \right )
 + \varepsilon_{ijk}
 S_{,jl}^{(2)} S_{,lm}^{(1)} \left ( \ddot{S}_{,km}^{(1)}
+ 2 \frac{\dot{a}}{a} \dot{S}_{,km}^{(1)} \right )
 = 0 .
\end{eqnarray}
By using first-, second-, and third-order equations
(\ref{eqn:1st-L-eq})--(\ref{eqn:1st-L-t-eq}),
(\ref{eqn:2nd-L-ts})--(\ref{eqn:2nd-L-spa-eq}),
(\ref{eqn:3rd-L-t-eqA})--(\ref{eqn:3rd-L-spa-eqB}), respectively, and
(\ref{eqn:3rd-T-ts}),
the equation is rewritten as
\begin{eqnarray}
&& \varepsilon_{ijk} \left (\ddot{s}_{k,j}^{T (4)} + 2 \frac{\dot{a}}{a}
 \dot{s}_{k,j}^{T (4)} \right ) \nonumber \\
&=& 4 \pi G \rho_b \varepsilon_{ijk} \times
 \left [ g_1 \left \{ g_{3b} - 2 g_1 (g_2-g_1^2) \right \}
  \psi_{,jl}^{(1)} \psi_{,kl}^{(3a)}
 + g_1 ( g_{3b} - 2 g_1^3 ) \psi_{,jl}^{(1)} \psi_{,kl}^{(3b)}
 \right . \nonumber \\
&& ~~~~~~~~~~~~~~ + g_1^4 \psi_{,jl}^{(1)} \zeta_{k,l}^{(3)}
 + g_1 g_{3a} \psi_{,kl}^{(1)} \psi_{,jl}^{(3a)}
 + g_1 g_{3b} \psi_{,kl}^{(1)} \psi_{,jl}^{(3b)}
 \nonumber \\
&& \left . ~~~~~~~~~~~~~~
 + g_1 g_{3T} \psi_{,kl}^{(1)} \zeta_{j,l}^{(3)}
 - g_1^4 \psi_{,jl}^{(1)} \psi_{,lm}^{(1)} \psi_{,km}^{(2)}
 \right ] .
\end{eqnarray}
Upon decomposition of the fourth-order perturbation by
Eq.~(\ref{eqn:4th-T-decompose}),
we obtain Eqs. (\ref{eqn:4th-T-g4a})--(\ref{eqn:4th-T-psi4d}).

\section{Higher-order loop correction for power spectrum}\label{sec:loop-p}

The generic formula about the power spectrum with $N$-loop corrections
in both real space and redshift space has been derived~\cite{{Okamura11}}. 
Here we suppose Gaussian initial fluctuation.
In real space, the power spectrum
is described as 
\begin{eqnarray} \label{eqn:loop-Pk}
P(k) &=& \exp \left [ -2 \sum_{n=1}^{\infty}
 \frac{k_{i_1} \cdots k_{i_{2n}}}{(2n)!} A_{i_1 \cdots i_{2n}}^{(2n)}
  \right ]
  \left \{ P_L (k) + \sum_{m=1}^{\infty}
  P_{\mbox{SPT}}^{m-{\mbox{loop}}} (k) \right \} \nonumber  \\
  && \times 
  \left [ 1+ \sum_{\ell=1}^{\infty} \frac{1}{\ell!} 
   \left ( 2 \sum_{n=1}^{\infty}
 \frac{k_{i_1} \cdots k_{i_{2n}}}{(2n)!} A_{i_1 \cdots i_{2n}}^{(2n)}
 \right )^{\ell} \right ] \,,
\end{eqnarray}
where
\begin{equation}
A_{i_1 \cdots i_{2n}}^{(2n)} =
 \left [ \prod_{i=1}^{2n} \int \frac{{\rm d}^3 p_i}{(2\pi)^3}
  \right ]
  \delta_D^3 \left (\sum_{i=1}^{2n} \bm{p}_i \right )
   C_{i_1 \cdots i_{2n}} (\bm{p}_1, \cdots , \bm{p}_{2n} ) \,.
\end{equation}
$\delta_D$ is Dirac's delta function.
$C_{i_1 \cdots i_{2n}}$ is a $2n-$order cumulant of the 
displacement field in Fourier space $\widehat{\bm{\Psi}}$:
\begin{equation}
\left < \widehat{\Psi}_{i_1} (\bm{p}_1)
 \cdots \widehat{\Psi}_{i_{2n}} (\bm{p}_{2n}) \right >_c
  = (-1)^{n-1} (2 \pi)^3 \delta_D^3
   \left (\sum_{i=1}^{2n} \bm{p}_i \right )
   C_{i_1 \cdots i_{2n}} (\bm{p}_1, \cdots , \bm{p}_{2n} ) \,.
\end{equation}
For derivation of the power spectrum, $N$-loop contribution
to the power spectrum predicted by SPT is required.
In $N$-loop correction of Lagrangian resumption,
the series in the exponent is expanded up to
$\mathcal{O} \left ( (P_L)^N \right)$. The remaining factor
is expanded up to $\mathcal{O} \left ( (P_L)^{N+1} \right)$. 
 
%
%
%

Following the formula~(\ref{eqn:loop-Pk}), three-loop correction
for the power spectrum is derived.
Here we extend correction terms.
\begin{eqnarray}
A_{ij}^{(2)} &=& \mathcal{A}_{ij}^{(11)} + \mathcal{A}_{ij}^{(22)}
 + 2 \mathcal{A}_{ij}^{(13)} + \mathcal{A}_{ij}^{(33)}
 + 2 \mathcal{A}_{ij}^{(24)} + 2 \mathcal{A}_{ij}^{(15)} +
  \cdots \,, \\
A_{ijlm}^{(4)} &=& 6 \mathcal{A}_{ijlm}^{(1122)}
 + 4 \mathcal{A}_{ijlm}^{(1113)} + \cdots \,.
\end{eqnarray}
The superscripts of $\mathcal{A}^{(\alpha \beta)}$ mean
that $\mathcal{A}$ consists of $\alpha$-order and
$\beta$-order perturbative quantities. 
In Gaussian initial fluctuation, $\mathcal{A}^{(111111)}$
does not appear.
The explicit form of three-loop correction
for the power spectrum is shown as follows:
\begin{eqnarray}
P(k) &=& \exp \left [-\frac{k^2}{6 \pi^2} \left (\mathcal{A}^{(11)}
 +\mathcal{A}^{(22)} + 2 \mathcal{A}^{(13)} + \mathcal{A}^{(33)} + 2\mathcal{A}^{(24)} + 2\mathcal{A}^{(15)}
  \right ) \right . \nonumber \\
   && ~~~~~~~~~ \left . -\frac{k_i k_j k_l k_m}{6}
   \left ( 3 \mathcal{A}_{ijlm}^{(1122)} + 2 \mathcal{A}_{ijlm}^{(1113)} \right ) \right ]
    \nonumber \\
 && \times \left [ P_L(k) + P_{\mbox{SPT}}^{1-{\mbox{loop}}} (k)
   + P_{\mbox{SPT}}^{2-{\mbox{loop}}} (k)
   + P_{\mbox{SPT}}^{3-{\mbox{loop}}} (k)  \right . \nonumber \\
   && ~~~~ 
   + P_L(k) \left \{ \frac{k^2}{6\pi^2}
    \left  (\mathcal{A}^{(11)} + \mathcal{A}^{(22)} + 2 \mathcal{A}^{(13)} + \mathcal{A}^{(33)}
     + 2\mathcal{A}^{(24)} + 2\mathcal{A}^{(15)} \right )  \right . \nonumber \\
    && ~~~~~~~~~~~~~
    + \frac{k^4}{72 \pi^4} (\mathcal{A}^{(11)})^2 + 2\mathcal{A}^{(11)} \mathcal{A}^{(22)}
      + 2 \mathcal{A}^{(11)} \mathcal{A}^{(13)} ) 
           + \frac{k^6}{1296 \pi^6} (\mathcal{A}^{(11)})^3 \nonumber \\
   && ~~~~~~~~~~~~~ \left . 
     + \frac{k_i k_j k_l k_m}{6}
   \left ( 3 \mathcal{A}_{ijlm}^{(1122)} + 2 \mathcal{A}_{ijlm}^{(1113)} \right )  
      \right \}  \nonumber \\
     && ~~~~~~ + \frac{k^2}{6\pi^2}
       P_{\mbox{SPT}}^{1-{\mbox{loop}}} (k) (\mathcal{A}^{(11)} + \mathcal{A}^{(22)} )
        + \frac{k^4}{72 \pi^4} P_{\mbox{SPT}}^{1-{\mbox{loop}}} (k)
        (\mathcal{A}^{(11)})^2 \nonumber \\
     && ~~~~~~ \left . + \frac{k^2}{6\pi^2}
       P_{\mbox{SPT}}^{2-{\mbox{loop}}} (k) \mathcal{A}^{(11)} \right ] \,.
\end{eqnarray}
For the computation for $A^{(15)}$, 5LPT solutions are required.
Therefore, three-loop correction for the power spectrum,
4LPT is not enough.

\end{document}